\documentclass{PoS}
\usepackage{epsfig}
\usepackage{bm}
\usepackage{amssymb}
\usepackage{fontenc}
\usepackage{times}
\usepackage{mathptmx}
\usepackage{graphicx}

\title{Exploring Complex-Langevin Methods for Finite-Density QCD}

\ShortTitle{Exploring Complex-Langevin Methods for Finite-Density QCD}

\author{\speaker{D.~K.~Sinclair}%
         \thanks{This research was supported in part by US Department of Energy
         contract DE-AC02-06CH11357}
         \\
HEP Division, Argonne National Laboratory, 9700 South Cass Avenue, Argonne, 
Illinois 60439, USA\\
E-mail: \email{dks@hep.anl.gov}}

\author{J.~B.~Kogut\\
Department of Energy, Division of High Energy Physics, Washington, DC 20585,
USA\\
and\\
Department of Physics -- TQHN, University of Maryland, 82 Regents Drive, 
College Park, MD 20742, USA\\
        E-mail: \email{jbkogut@umd.edu}}

\abstract{QCD at non-zero chemical potential ($\mu$) for quark number has a
complex fermion determinant and thus standard simulation methods for lattice
QCD cannot be applied. We therefore simulate this theory using the
Complex-Langevin algorithm with Gauge Cooling in addition to adaptive methods,
to prevent runaway behaviour. Simulations are performed at zero temperature on
a $12^4$ lattice with 2 quarks which are light enough that $m_N/3$ is
significantly larger than $m_\pi/2$. Preliminary results are qualitatively as
expected. The quark-number density is close to zero for $\mu < m_N/3$, beyond
which it increases, eventually reaching its saturation value of $3$ for $\mu$
sufficiently large. The chiral condensate decreases as $\mu$ is increased 
approaching zero at saturation, while the plaquette increases towards its 
quenched value. We have yet to observe the transition to nuclear matter at
$\mu \approx m_N/3$, presumably because the runs for $\mu$ between $m_N/3$ and
saturation have yet to equilibrate.}

\FullConference{The 33rd International Symposium on Lattice Field Theory\\
                 14 -18 July  2015\\
                 Kobe International Conference Center, Kobe, Japan}

\begin{document}

\section{Introduction}

QCD at a finite chemical potential $\mu$ for quark number has a complex action
which prevents the direct application of simulation methods based on
importance sampling. The Langevin equation is a stochastic differential
equation for the evolution of the classical fields in a fictitious time, which
does not rely on importance sampling. It is, in fact, a special case of the
hybrid molecular-dynamics algorithm, where each trajectory consists of a
single update.

The Langevin equation can be extended to complex actions by complexifying the
fields \cite{Parisi:1984cs,Klauder:1983nn,Klauder:1983zm,Klauder:1983sp}. 
In the case of QCD this means promoting the gauge fields from 
$SU(3)$ to $SL(3,C)$. Unfortunately, there is no proof that the long-time
evolution of the fields under this Complex Langevin equation (CLE) provides a
limiting value for observables. Even when this process does converge, the
values it provides for observables are not guaranteed to be correct.

After successfully applying the CLE to spin models (see for example
\cite{Karsch:1985cb}), people were encouraged to apply it to lattice QCD at
finite $\mu$. Early attempts at applying the CLE to QCD were stymied by
runaway behaviour, which was not corrected by adaptive methods. Recently it
has been noted that at least some of this undesirable behaviour is due to the
production of unbounded gauge transformations of compact gauge fields. Such
behaviour can be controlled by gauge transforming to a gauge which minimizes
the magnitudes of the gauge fields and hence their distance from the $SU(3)$ 
manifold\cite{Seiler:2012wz}. This is called Gauge Cooling.

This has revived interest in the CLE for QCD at finite $\mu$. These methods 
have been tested on simple models and for quark masses large enough that
hopping-parameter methods can be applied \cite{Aarts:2008rr,Aarts:2008wh,
Aarts:2011zn,Aarts:2010gr,Aarts:2014kja,Aarts:2014bwa}.
In addition, studies have been made of
the conditions under which the CLE converges to the correct results 
\cite{Aarts:2011ax,Aarts:2013uxa,Aarts:2012ft,Aarts:2013uza,Nishimura:2015pba,
Nagata:2015uga,Makino:2015ooa}. QCD at
finite $\mu$ and small masses has been simulated and the results compared with
the heavy quark methods, for larger masses \cite{Sexty:2013ica}. 
Very recently this has been extended
to larger lattices at finite temperatures, where the transition from 
hadron/nuclear matter to a quark-gluon plasma is observed and results are 
compared with those from reweighting methods \cite{Fodor:2015doa}.

We are simulating QCD at zero temperature and finite $\mu$ for light quarks
using the CLE, to test directly if it converges and produces believable results.
We present preliminary results of our explorations.

\section{Complex Langevin for finite density Lattice QCD}

If $S(U)$ is the gauge action after integrating out the quark fields, the
Langevin equation for the evolution of the gauge fields $U$ in Langevin 
time $t$ is:
\begin{equation}
-i \left(\frac{d}{dt}U_l\right)U_l^{-1} = -i \frac{\delta}{\delta U_l}S(U)
+\eta_l
\end{equation}
where $l$ labels the links of the lattice, and 
$\eta_l=\eta^a_l\lambda^a$. Here $\lambda_a$ are the Gell-Mann 
matrices for $SU(3)$. $\eta^a_l(t)$ are Gaussian-distributed random 
numbers normalized so that:
\begin{equation}
\langle\eta^a_l(t)\eta^b_{l'}(t')\rangle=\delta^{ab}\delta_{ll'}\delta(t-t')
\end{equation}

The complex-Langevin equation has the same form except that the $U$s are now
in $SL(3,C)$. $S$, now $S(U,\mu)$ is 
\begin{equation}
S(U,\mu) = \beta\sum_{_{\bm\Box}} \left\{1-\frac{1}{6}{\rm Tr}[UUUU+(UUUU)^{-1}]
\right\} - \frac{N_f}{4}{\rm Tr}\{\ln[M(U,\mu)]\}
\end{equation}
where $M(U,\mu)$ is the staggered Dirac operator. Note: backward links
are represented by $U^{-1}$ not $U^\dag$. Note also that we have 
chosen to keep the noise-vector $\eta$ real. $\eta$ is gauge-covariant under
$SU(3)$, but not under $SL(3,C)$. This means that gauge-cooling is non-trivial.
Reference~\cite{Nagata:2015uga} indicates why this is not expected to change
the physics. After taking $-i\delta S(U,\mu)/\delta U_l$, the cyclic
properties of the trace are used to rearrange the fermion term so that it
remains real for $\mu=0$ even after replacing the trace by a stochastic
estimator.

To simulate the time evolution of the gauge fields we use the partial 
second-order formalism of Fukugita, Oyanagi and Ukawa.
\cite{Ukawa:1985hr,Fukugita:1986tg,Fukugita:1988qs} 

After each update, we
gauge-fix iteratively to a gauge which minimizes the unitarity norm
-- gauge cooling \cite{Seiler:2012wz}:
\begin{equation}
F(U) = \frac{1}{4V}\sum_l{\rm Tr}\left[U_l^\dag U_l + (U_l^\dag U_l)^{-1} 
     - 2\right] \ge 0,
\end{equation}
where $V$ is the space-time volume of the lattice.

\section{Zero-temperature simulations}

\subsection{$\mu=0$}

For $\mu=0$ and infinite precision, Complex Langevin becomes Real Langevin.
At 64-bit precision, roundoff allows the gauge fields to move (slowly) off the
$SU(3)$ manifold. For $\beta=5.2$, $m=0.05$ on an $8^4$ lattice we observe 
runaway solutions, even after Gauge Cooling!

\begin{figure}[htb]
\parbox{2.9in}{
\epsfxsize=2.9in
\centerline{\epsffile{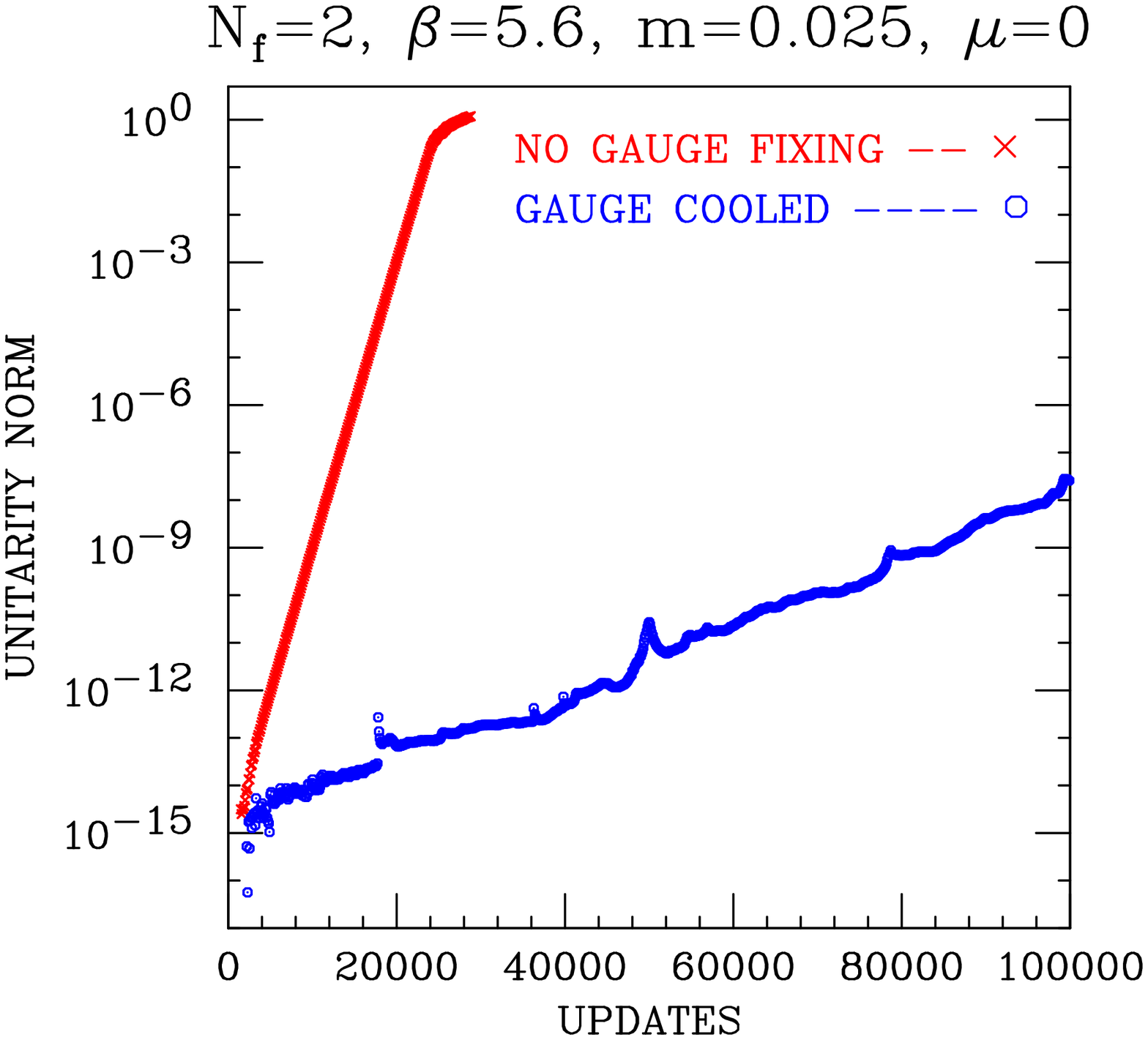}}
\caption{Unitarity norms for runs on a $12^4$ lattice. Red curve is for
run without gauge cooling. Blue curve is for run with 10-step gauge cooling.}
\label{fig:unorm0}
}
\parbox{0.2in}{}
\parbox{2.9in}{
\epsfxsize=2.9in 
\centerline{\epsffile{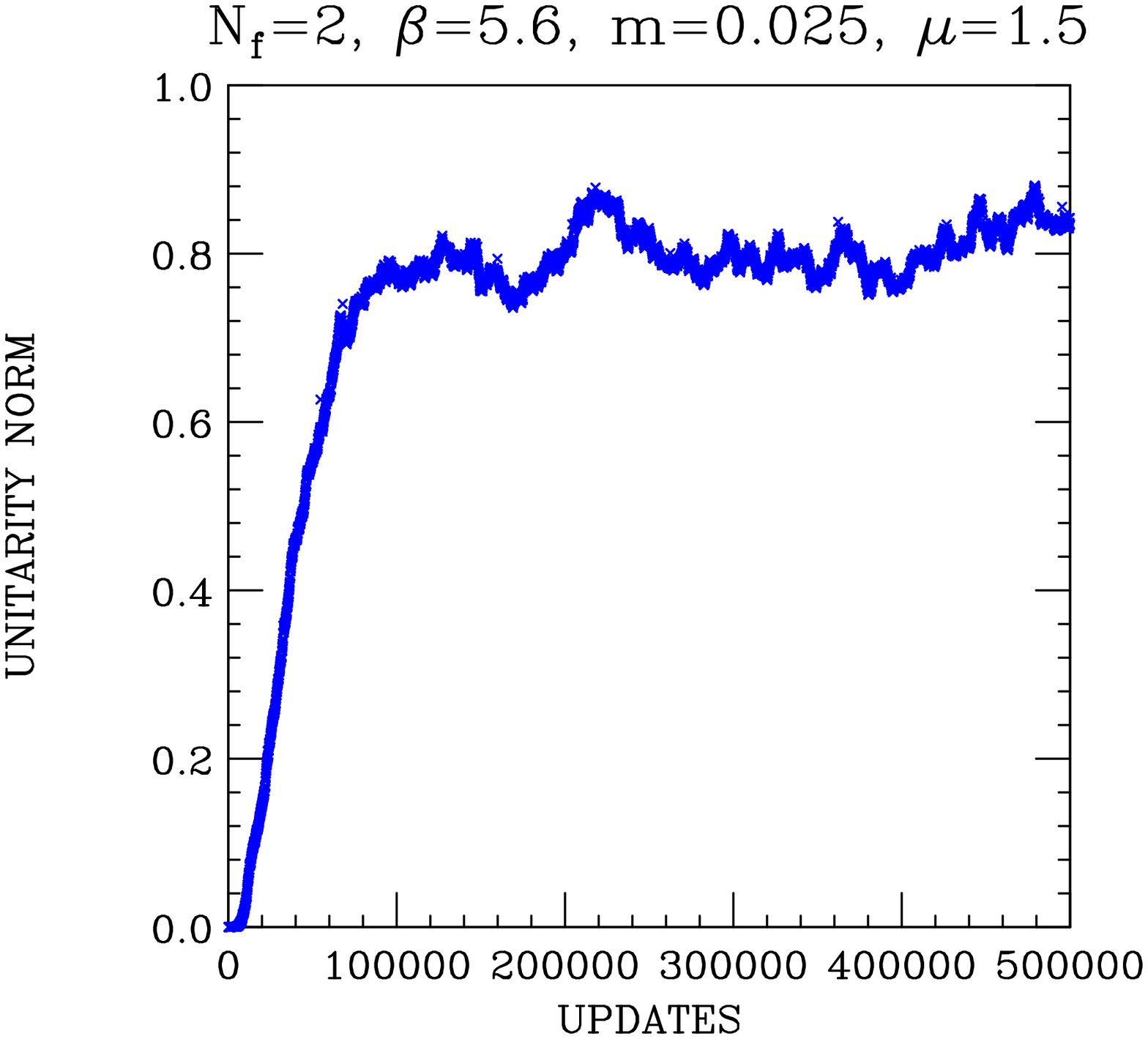}}
\caption{Unitarity norms for run at $\mu=1.5$ on a $12^4$ lattice with 5-step
gauge-cooling.}
\label{fig:unorm1.5}
}
\end{figure}

For $\beta=5.6$, $m=0.025$ on a $12^4$ lattice without gauge cooling, we
observe runaway solutions. With gauge cooling, the trajectory moves slowly off
the $SU(3)$ manifold. We perform 100,000 updates with input $dt=0.01$. 
Adaptively rescaling to keep the drift(force) term under control $dt$ is
reduced to $dt_{\it adaptive} \approx 0.00108$, so the total run takes
$\approx 108$~Langevin time units, at the end of which the unitarity norm%
~$\approx 2.5 \times 10^{-8}$. This we can probably tolerate, especially since
we expect it to improve with weaker couplings and larger lattices.
Figure~\ref{fig:unorm0} shows the time evolution of the unitarity norm with and
without gauge-cooling.

For this run plaquette~$=0.4351(1)$, whereas the RHMC algorithm
gives~$0.43588(4)$, and $\langle\bar{\psi}\psi\rangle=0.208(2)$,
RHMC~$0.2142(8)$. This is reasonable agreement for a short run with an inexact
algorithm.

\subsection{$\mu \ne 0$}

We simulate on a $12^4$ lattice at $\beta=5.6$, $m=0.025$ with $\mu > 0$.
Potentially important $\mu$ values include $m_\pi/2 \approx 0.21$ and
$m_N/3 \approx 0.33$ (masses from HEMCGC collaboration 
\cite{Bitar:1990cb,Bitar:1993rk,Bitar:1990wk}). 
The first is the position of the expected transition for the phase-quenched
approximation. The second is the approximate position of the expected
transition to nuclear matter. We start with a limited number of $\mu$ values
to probe the various regions of the zero-temperature phase diagram. The values
we choose are $0.1$, $0.2$, $0.25$, $0.35$, $0.5$, $0.9$ and $1.5$. In each
case we start the simulation from an ordered start and use 5-step
gauge-cooling.

The first thing we look for, is evidence that the trajectories for a given set
of parameters are restricted to a compact region of the $SL(3,C)$ manifold.
Without this it is (almost) impossible for these simulations to produce
meaningful results. If the simulations do converge to a limiting distribution,
one must then address the question as to whether this is the correct limit.

At $\mu=1.5$ we have performed sufficient updates for the unitarity norm to
level off. The unitarity norm appears to have leveled off, indicating that the
system is evolving over a compact domain of $SL(3,C)^{4V}$ The evolution
of this norm over the trajectory is shown in figure~\ref{fig:unorm1.5}. The
quark number density $j_0=2.9998(2)$. Hence the system has reached saturation
where $j_0=3$, as expected for large $\mu$. This is where each site is occupied
with 3 quarks in a colour singlet state (nucleon). The chiral condensate
$\langle\bar{\psi}\psi\rangle=0.5(1.4) \times 10^{-5}$ -- small and consistent
with zero as expected. The plaquette $P=0.4679(2)$, consistent with the idea
that, at saturation, the quarks are frozen out and the system approximates
quenched QCD. The quenched plaquette at $\beta=5.6$, $P=0.47553(2)$.
The total trajectory length $\approx 46$ time-units. In fact, after 
equilibration, $dt_{\it adaptive} \approx 0.000066$.

To date, of the other $\mu$ values we are simulating, $\mu=0.1$ and $\mu=0.2$
appear to have equilibrated. $\mu=0.25$ appears to be close to equilibrating.
$\mu=0.35$, $\mu=0.5$ and $\mu=0.9$ have yet to equilibrate. We believe that
this is because these runs need more updates, since their unitarity norms have
yet to reach values achieved for the $\mu=0.1$ simulations.

\begin{figure}[hbt]
\parbox{2.9in}{
\epsfxsize=2.9in
\centerline{\epsffile{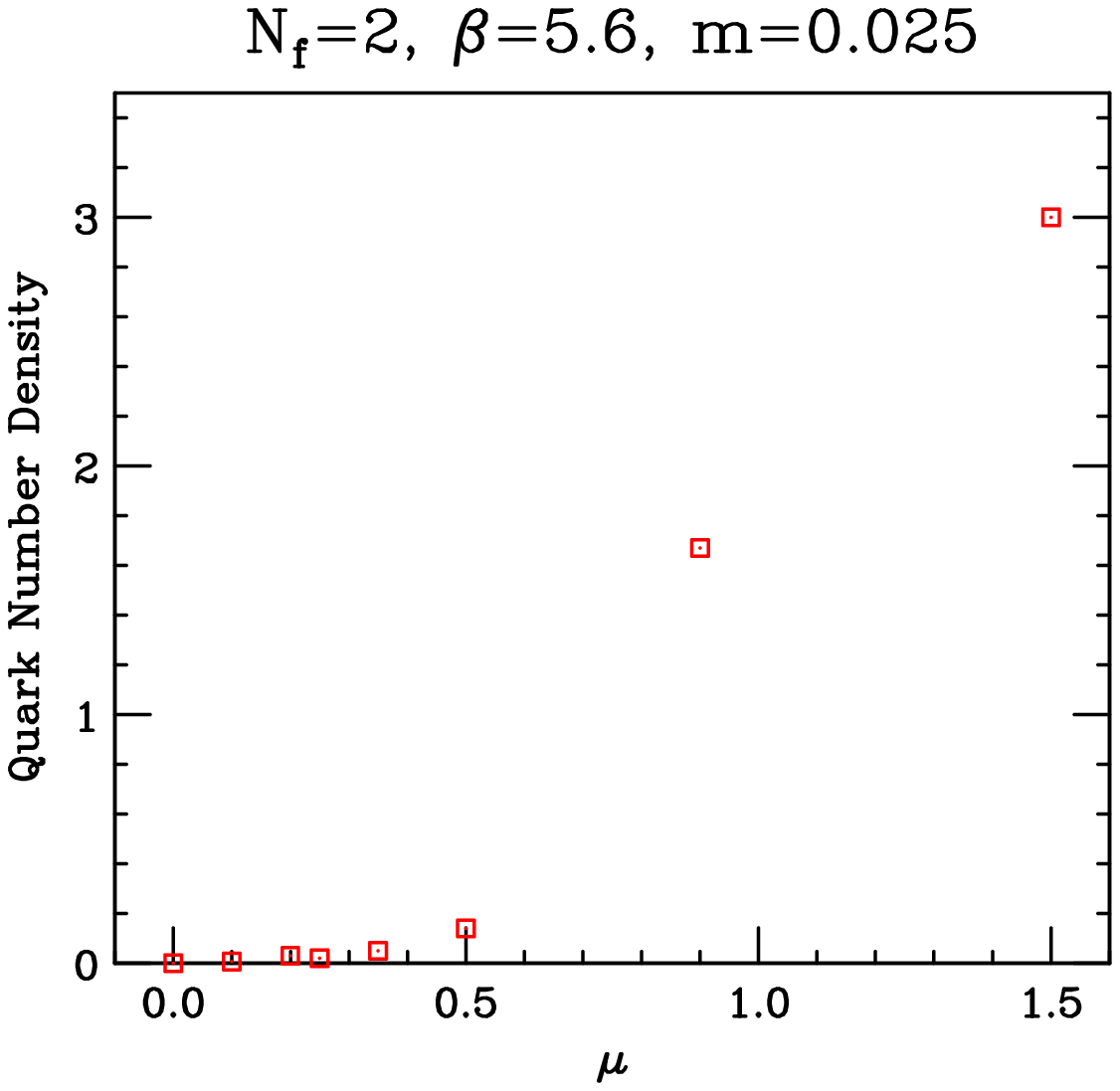}}
\caption{Quark number density, normalized to one staggered quark (4-flavours), 
as a function of $\mu$. Errors not known.}
\label{fig:qnd}
}
\parbox{0.2in}{}
\parbox{2.9in}{
\epsfxsize=2.9in
\centerline{\epsffile{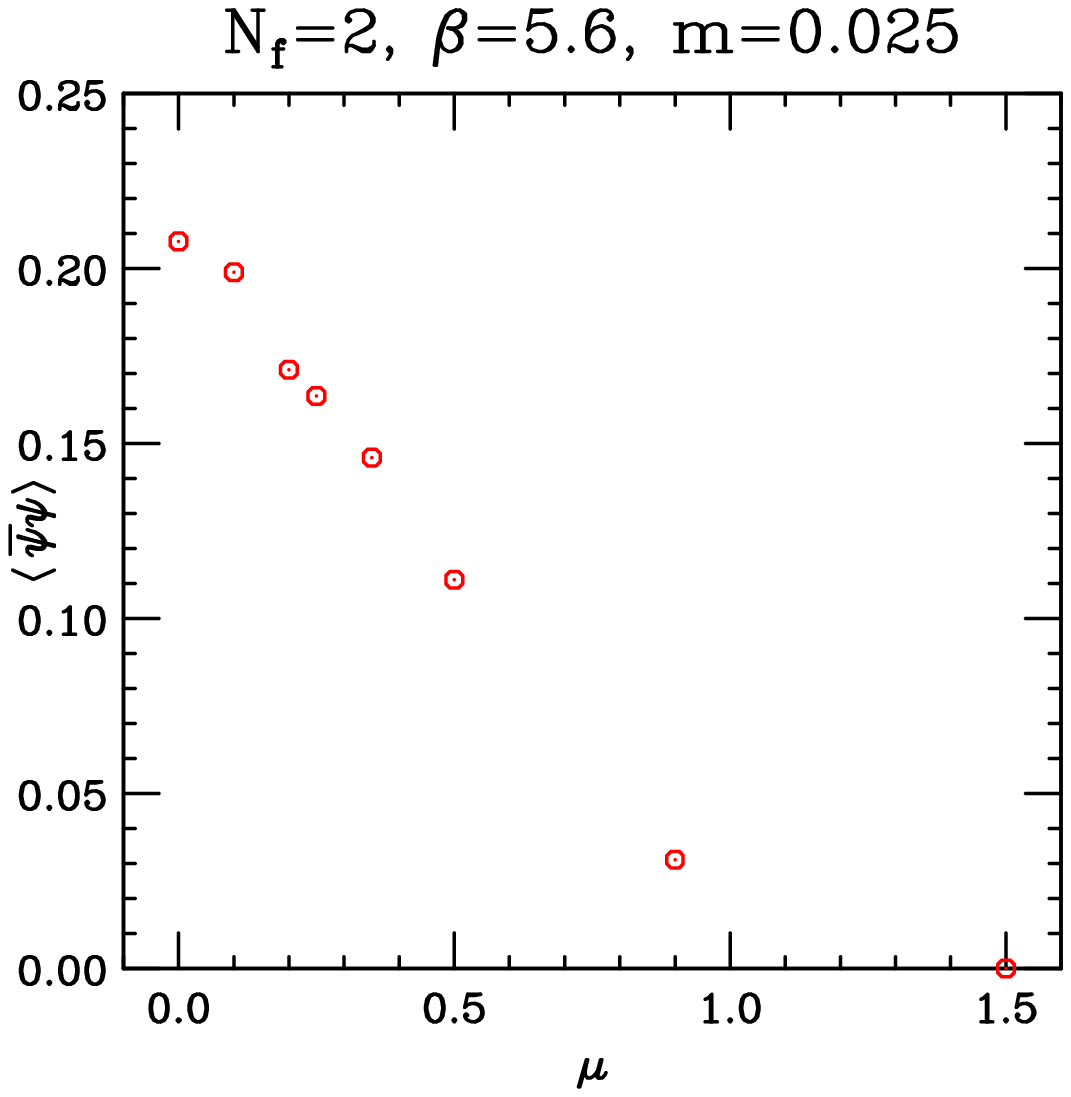}}
\caption{Chiral condensate, normalized to one staggered quark (4-flavours), as 
a function of $\mu$. Errors not known.}
\label{fig:pbp} 
}                                                             
\end{figure}        

We present `data' for the quark-number densities (figure~\ref{fig:qnd}), and
chiral condensates (figure~\ref{fig:pbp}) as functions of $\mu$ with the
understanding that the points at $\mu=0.35$, $0.5$ and $0.9$ are expected to
change to become closer to the values at $\mu=1.5$ as the system equilibrates.
This is because, since we start each run with all gauge links on the $SU(3)$
manifold, then as the system equilibrates, the gauge links move away from this
$SU(3)$ manifold. Because not all $\mu$s are equilibrated, we have not included
error-bars in these figures.

These preliminary results (each point represents 90,000~--~500,000 sweeps/%
updates of the lattice) agree qualitatively with our expectations. The 
quark-number density remains close to zero for $\mu < m_N/3$. For larger $\mu$s
it becomes non-zero, increasing towards its saturation value of $3$ as $\mu$ is
increased. The chiral condensate decreases monotonically from its $\mu=0$ value
as $\mu$ is increased, approaching zero at saturation. We will need to wait
until each point has equilibrated to where it is clear that the gauge fields
are varying over a compact region in the $SL(3,C)$ manifold, before we can get
truly quantitative results. This takes longer for $\mu > m_N/3$ since it takes
more iterations to invert the Dirac operator as $\mu$ increases, until close
to saturation, and $dt_{\it adaptive}$ is smaller. Since each run is starting
from the $SU(3)$ manifold, we expect metastability for $\mu > m_N/3$ due to
the presence of a supposed first-order transition at $\mu \approx m_N/3$. This
will also slow down the approach to equilibrium just above the transition.

Because the runs for $\mu$ just above the transition have yet to equilibrate,
we have been unable to observe this transition to nuclear matter.

\section{Summary, discussion and outlook}

We apply Complex-Langevin simulations with gauge cooling to lattice QCD at
finite quark-number chemical-potential ($\mu$) at zero temperature. Our
current simulations are on a $12^4$ lattice with $N_f=2$, $\beta=5.6$,
$m=0.025$. Preliminary results look promising, but more simulations are
needed. Adaptive updating with gauge cooling does appear to stabilize the
algorithm. However, this only appears to work provided the gauge coupling is
not too strong. Since gauge cooling does not completely fix the gauge -- 
the unitarity norm is invariant under $SU(3)$ gauge transformations -- it is
possible that further gauge fixing might improve the situation. Fixing to 
Landau gauge in the $SU(3)$ subgroup suggests itself. 

We need answers to the following.
Do these simulations converge and converge to the correct limit?
Do we observe a phase transition to nuclear matter at $\mu \approx m_N/3$?
Is there a spurious transition at $\mu \approx m_\pi/2$?
Do these simulations produce the expected 2-flavour colour-superconductor at
large $\mu$ ($\mu > m_N/3$)?.

Smaller masses are needed -- $m=0.01$(?). We also need larger lattices, weaker
coupling... Our current code is inefficient serial code which needs
improvement. When we are convinced that the algorithm works we will
parallelize our code.

We will also investigate whether we can make a fully second-order version. In 
addition we will investigate whether it makes sense to make a complex extension
of hybrid molecular-dynamics. Would complex hybrid molecular-dynamics be 
expected to converge to the correct limit, if the complex langevin does? If so,
such an algorithm would expect to be faster and have smaller errors than the
complex-langevin methods.

A precise measurement of the value of $\mu$ at the transition to nuclear
matter ($\mu_c$) would yield the binding energy/nucleon ($\epsilon_b$) in the
absence of electromagnetic interactions, since $\mu_c=(m_N-\epsilon_b)/3$.
However, since $\epsilon_b < 2\%\, m_N$, this will be a formidable task. More
accessible nuclear physics will be to study the propagation of hadrons in the
nuclear-matter medium. If our 2-flavour simulations are successful, we will
also simulate $N_f=3$ and try to observe the 3-flavour colour-superconductor
with its colour-flavour locking.

We also plan to simulate at high temperatures, near to the finite-temperature
phase transition, and look for the critical endpoint. We will try to determine
how good is the resonance-gas model. Also planned are investigations of the
phase structure of $(2+1)$-flavour QCD with independent chemical potentials
for the $(u,d)$ and $s$ quarks.

\section*{Acknowledgements}

These simulations are being performed on PCs belonging to Argonne's HEP 
Division, Edison and Carver at NERSC, and Blues at LCRC, Argonne.

\end{document}